\documentclass[fleqn,10pt]{wlscirep}
\usepackage[utf8]{inputenc}
\usepackage[T1]{fontenc}

\usepackage{amsmath}  % For mathematical commands
\usepackage{amsfonts} % For additional font support
\usepackage{amssymb} 
\usepackage{bbm}
\usepackage{cite}     % For handling citations
\usepackage{xcolor}
\usepackage{subcaption}
\usepackage{pgfplots} % Required for the axis environment
\pgfplotsset{compat=1.18}
\usepackage{tikz}
\usetikzlibrary{shapes.geometric, arrows, fit, positioning, calc}
\usepackage{arydshln} % Add this in the preamble for dotted lines
\title{Quantum Annealing Feature Selection on Light-weight Medical Image Datasets}

\author[1,*]{Merlin A. Nau}
\author[2, 3]{Luca A. Nutricati}
\author[4]{Bruno Camino}
\author[2,5]{Paul A. Warburton}
\author[1]{Andreas K. Maier}
\affil[1]{Pattern Recognition Lab, Friedrich-Alexander-Universität Erlangen-Nürnberg, Erlangen, 91052, Germany}
\affil[2]{London Centre of Nanotechnology, University College London, London, WC1H 0AH, United Kingdom}
\affil[3]{Rudolf Peierls Centre for Theoretical Physics, University of Oxford, Oxford, OX1 3PU, United Kingdom}
\affil[4]{Department of Chemistry, University College London, London, WC1H 0AJ, United Kingdom}
\affil[5]{Department of Electronic \& Electrical Engineering, University College London, London, WC1E 7JE, United Kingdom}

\affil[*]{merlin.nau@fau.de}

\keywords{Medical Imaging, Machine Learning, Quantum Computing, Quantum Annealing, Image Reconstruction}
\begin{abstract}
We investigate the use of quantum computing algorithms on real quantum hardware to tackle the computationally intensive task of feature selection for light-weight medical image datasets. Feature selection is often formulated as a $k$ of $n$ selection problem, where the complexity grows binomially with increasing $k$ and $n$. As problem sizes grow, classical approaches struggle to scale efficiently. Quantum computers, particularly quantum annealers, are well-suited for such problems, offering potential advantages in specific formulations. We present a method to solve larger feature selection instances than previously presented on commercial quantum annealers. Our approach combines a linear Ising penalty mechanism with subsampling and thresholding techniques to enhance scalability. The method is tested in a toy problem where feature selection identifies pixel masks used to reconstruct small-scale medical images. The results indicate that quantum annealing-based feature selection is effective for this simplified use case, demonstrating its potential in high-dimensional optimization tasks. However, its applicability to broader, real-world problems remains uncertain, given the current limitations of quantum computing hardware.
\end{abstract}
\begin{document}

\flushbottom
\maketitle
\thispagestyle{empty}

\section*{Introduction}

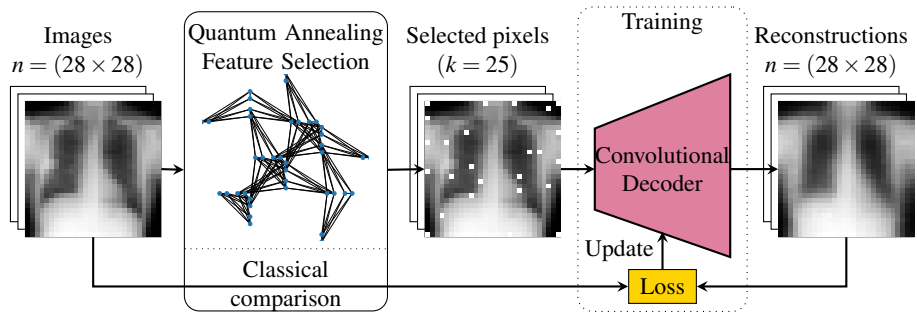
\begin{figure}[ht]
    \centering
    \begin{tikzpicture}[node distance=1cm, scale=0.9, every node/.style={scale=0.9}]

    % Styles
    \tikzstyle{block} = [rectangle, minimum width=1.0cm, minimum height=0.5cm, text centered, draw=black, fill=blue!30]
    \tikzstyle{arrow} = [thick,->,>=stealth]
    \tikzstyle{imageblock} = [rectangle, minimum width=2.0cm, minimum height=2.0cm, draw=black, inner sep=0]
    \tikzstyle{featureselection} = [rectangle, rounded corners=0.2cm, text width=3.0cm, text height=4.0cm, draw=black, inner sep=0]
    \tikzstyle{training} = [rectangle, rounded corners=0.2cm, minimum width=2.5cm, minimum height=4.5cm, draw=black, dotted, inner sep=0]

    % Dataset Block
    % Backmost rectangle
    % Backmost rectangle

    % First rectangle not shifted
    \node (rect1) [imageblock, fill=white, label={[align=center]above: Images \\$n = (28 \times 28)$}] {};
    \node (rect2) [imageblock, fill=white, below right=0.1cm and 0.1cm of rect1.north west]{};
    \node (image) [imageblock, fill=white, below right=0.1cm and 0.1cm of rect2.north west] {
        \includegraphics[width=2.0cm,height=2.0cm]{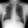}
    };

    \node (featureselection) [featureselection, below right= -1.9cm and 0.5cm of rect1.east] {
    % Two-line label with padding
    \centering
    \vspace{-4.0cm}\\
    Quantum Annealing\\Feature Selection \\ 
    % Padding between label and image
    \vspace{0.05cm}
    % Image
    \includegraphics[width=2.45cm, trim=80 80 80 80, clip]{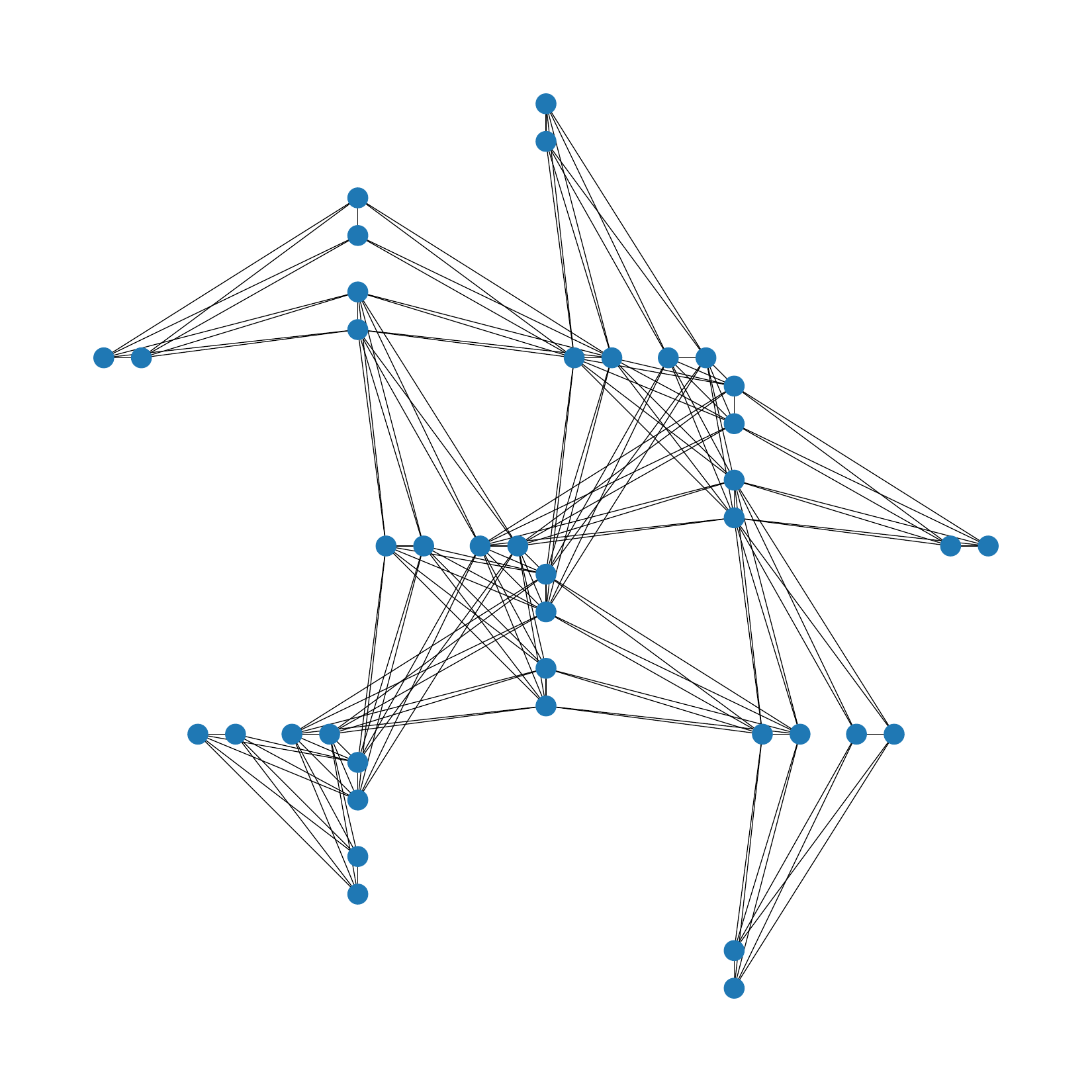}\\
    % Padding between image and line
    \vspace{-0.3cm}
    % Dotted line
    \tikz[baseline] \draw[black, dotted] (0,0) -- (3.0,0); \\ % Dotted line
    % Padding between line and text
    %\vspace{0.2cm}
    % Text below the line
    Classical comparison \\
    % Padding to ensure proper alignment
    \vspace{0.2cm}
};

% \node[anchor=center] at (2.5, -1.5)) {\includegraphics[width=2.5cm, trim=80 80 80 80, clip]{pegasus_topology.pdf}};

    \node (srect1) [imageblock, right=3.5cm of rect1, fill=white, label={[align=center]above: Selected pixels \\$(k=25)$}] {};
    \node (srect2) [imageblock, fill=white, below right=0.1cm and 0.1cm of srect1.north west]{};
    \node (selected_pixels) [imageblock, fill=white, below right=0.1cm and 0.1cm of srect2.north west] {
        \includegraphics[width=2.0cm,height=2.0cm]{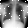}
    };

    % Funnel-like decoder
    \node (decoder) at ([xshift=1.5cm] selected_pixels.east) {};
    \draw[thick, fill=purple!50] 
      ([xshift=-1.0cm, yshift=-0.6cm] decoder.center) -- % Narrow left side
      ([xshift=-1.0cm, yshift=0.6cm] decoder.center) -- % Start of widening
      ([xshift=1.0cm, yshift=1.4cm] decoder.center) -- % Wider middle
      ([xshift=1.0cm, yshift=-1.3cm] decoder.center) -- % Bottom narrowing
      cycle;
    
    % Labels for decoder
    \node[align=center] at (decoder.center) {Convolutional \\ Decoder};

    \node (mse) [block, below=1.33cm of decoder.center, fill=yellow!70!orange] {Loss};

    \node (rrect1) [imageblock, right=4.7cm of srect1.west, fill=white, label={[align=center]above: Reconstructions \\$n=(28 \times 28)$}] {};
    \node (rrect2) [imageblock, fill=white, below right=0.1cm and 0.1cm of rrect1.north west]{};
    \node (reconstructed) [imageblock, fill=white, below right=0.1cm and 0.1cm of rrect2.north west] {
        \includegraphics[width=2.0cm,height=2.0cm]{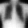}
    };

    \node (training) [training, above=-1.9cm of decoder.center, label={[align=center, yshift=4.55cm]below: Training}] {};

    % Arrows with labels
    %\draw [arrow] (image.east) -- (featureselection.east) {FS} (selected_pixels.west);
    \draw [arrow] (image.east) -- node[midway, above] {} ([yshift=-0.125cm] featureselection.west);
    \draw [arrow] ([yshift=-0.125cm] featureselection.east) -- node[midway, above] {} (selected_pixels.west);
    \draw [arrow] (selected_pixels.east) -- node[midway, above] {} ([xshift=-1.0cm] decoder.center);
    \draw [arrow] ([xshift=1.0cm] decoder.center) -- node[midway, above] {} (reconstructed.west);

    % Additional arrows
    \draw [arrow] (image.south) -- ++(0,0) |- node[midway, right] {} (mse.west);
    \draw [arrow] (reconstructed.south) -- ++(0,0) |- node[midway, left] {} (mse.east);
    \draw [arrow] (mse.north) -- node[midway, left] {Update} ([yshift=-0.8cm] decoder.south);

    \end{tikzpicture}
    \caption{Illustration of the feature selection process using quantum annealing, or classically, to extract pixels and train a convolutional decoder for reconstruction.}
    \label{fig:decoder}
\end{figure}

Medical imaging plays a pivotal role in modern medicine, offering essential insights for diagnosis and therapy. With the increasing complexity and sheer volume of imaging data, manual analysis by physicians has become increasingly challenging. To address these challenges, machine learning (ML), and in particular deep learning (DL), have proven to be transformative tools, excelling in tasks such as disease classification, segmentation, and regression~\cite{litjens2017survey, shen2017deep}. Using neural architectures to learn complex patterns from large datasets, these methods have significantly improved diagnostic efficiency and accuracy. In particular, the strength of neural networks lies in their ability to learn extracted features, which are then used for subsequent tasks. Since most ML and DL methods scale with the number of samples and the number of features, the process quickly becomes resource intensive. Feature selection (FS) reduces input dimensionality by retaining the most relevant features while minimizing redundancy. Therefore, FS has the potential to simplify ML models, improve interpretability, and reduce computational complexity, often enhancing robustness and generalization. FS methods are typically categorized into: \textit{filter methods}, which assess features using statistical metrics like mutual information (MI) or label correlation, and \textit{embedded methods}, which integrate FS into the model training process~\cite{chandrashekar2014survey}.

FS is a computationally intensive problem, as identifying the optimal subset from $n$ features involves evaluating $\binom{n}{k} = \frac{n!}{k!(n-k)!}$ subsets for $k$ features, resulting in binomial complexity as $n$ grows. To address this, FS can be reformulated as a quadratic unconstrained binary optimization (QUBO) problem, enabling efficient optimization techniques. This approach also opens the door to novel quantum computing methods, such as adiabatic quantum computing (AQC), which are well-suited for solving QUBO-based FS problems.

Quantum computing offers a novel paradigm for addressing computational bottlenecks in ML in the future, including FS. By harnessing quantum phenomena such as superposition and entanglement, quantum computers can in principle explore vast solution spaces more efficiently than classical counterparts~\cite{schuld2015introduction, biamonte2017quantum, kadowaki1998quantum, heim2015quantum, harris2018phase,king2018observation,albash2021comparing,king2021scaling}. Although current quantum hardware, known as noisy intermediate-scale quantum devices~\cite{preskill2018quantum}, is still limited in scale and precision, it holds immense potential for tasks requiring combinatorial optimization.

Adiabatic quantum computing is, in principle, highly effective for optimization problems like FS, operating by evolving a quantum system to an optimal solution~\cite{born1928beweis, farhi2001quantum, mcgeoch2014adiabatic}. Commercial AQC solutions, such as D-Wave's quantum annealers~\cite{LantingAQC2017}, provide practical platforms for solving QUBO problems, enabling exploration of quantum FS on real-world datasets like medical imaging.

Recently, experiments have been conducted with QUBO-based FS methods, executed on simulated and real quantum hardware~\cite{dwave2020feature, mucke2023feature}. The tutorial by D-Wave provides a starting point for MI-based feature selection using quantum annealing (QA) on toy problems~\cite{dwave2020feature}. In this manuscript, we follow Muecke et al.~\cite{mucke2023feature}, who introduced an MI-based QUBO for quantum FS on MNIST and small synthetic datasets. In particular, they investigated the quantum FS for classification, but also for lossy compression of images using an encoder trained on the selected features. For their MI-QUBO, they show classically solved results for up to $n=784$. However, when moving to current quantum hardware, the size constraints allow their algorithm to be applied on synthetic experiments with a maximum of $n=34$ features. QA based FS has also been tested on tasks such as classification of hyper spectral images~\cite{otgonbaatar2021quantum}, RNA sequencing analysis~\cite{romero2024quantum}, recommendation systems~\cite{nembrini2021feature}, or the selection of radiomic features~\cite{felefly2023explainable}, following the presented prior work. Aside from FS, QA has been actively investigated for use cases in medical imaging and computer vision, for example, tomographic image reconstruction~\cite{jun2023highly, nau2023exploring, nau2024improving}, segmentation~\cite{jun2023quantum}, and super resolution~\cite{choong2023quantum}.% and acquisition optimization~\cite{fuchs2023optimization, prjamkov2024comparison}.

In this work, we explore how current QA can be used for FS on larger problem instances by focusing on a light-weight example. We provide the reader with an introduction to QA and present our QA-based FS method. Our experiments are based on the MedMNIST data set~\cite{yang2023medmnist}, a benchmark for medical imaging consisting of standardized $28 \times 28$ pixel images across multiple modalities. For simplicity, we treat each of the 784 pixels as individual features, recognizing that pixels are not optimal feature descriptors. Our goal is not to compete against the performance of DL in medical imaging but rather to provide an accessible demonstration of how quantum technologies can be applied in ML, enabling problem sizes previously considered infeasible for current quantum devices.

\begin{figure}[!ht]
    \centering
    \begin{subfigure}[t]{0.35\textwidth}
        \centering
        \includegraphics[keepaspectratio, width=\textwidth, trim=65 65 65 65, clip]{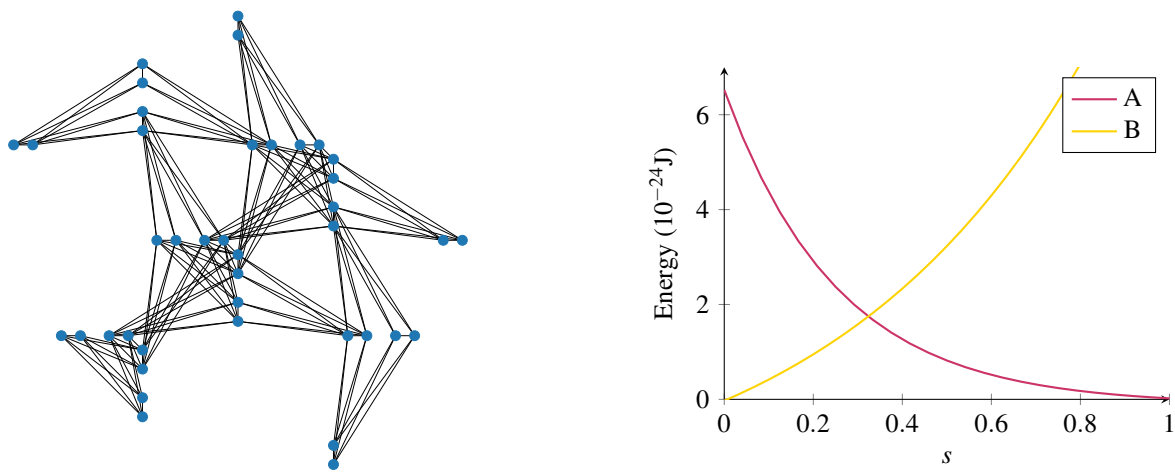}
    \end{subfigure}
    \hfill
    \begin{subfigure}[t]{0.63\textwidth}
        \centering
        \begin{tikzpicture}
            \begin{axis}[
                width=7.5cm, height=6.0cm,
                xlabel={$s$},
                ylabel={Energy ($10^{-24}$J)},
                legend pos=north east,
                axis lines=left,
                grid=none,
                xmin=0, xmax=1,
                ymin=0, ymax=7
            ]
                % Exponential decay function A
                \addplot[purple!80, thick, domain=0:1] {6.63 * exp(-3.93 * x) - 0.11};
                \addlegendentry{A}
    
                % Optional alternative for B (example)
                \addplot[yellow!70!orange, thick, domain=0:0.8] {2.35 * exp(1.74 * x) - 2.38}; 
                \addlegendentry{B}
            \end{axis}
        \end{tikzpicture}
    \end{subfigure}
    \caption{(Left) A part of the Pegasus topology implemented in the quantum annealing D-Wave \texttt{Advantage\_system4.1} architecture, where qubits (blue circles) are connected with couplings (black lines) to a maximum of 15 other qubits. (Right) Anneal schedule parameters, where $A(s)$ and $B(s)$ scale the transverse field and Ising contributions, respectively. These coefficients are functions of the parameter $s \in [0,1]$, which depends on the physical time $t$.}
    \label{fig:combined_figures}
\end{figure}

We present three key contributions. First, we implement QA-based FS on six lightweight medical imaging datasets. Second, we address hardware limitations by employing subsampling, thresholding, and a "linear Ising penalty" trick~\cite{ohzeki2020breaking, mirkarimi2024quantum}, enabling execution on current commercially available quantum annealers. Finally, we demonstrate the utility of the selected features through a convolutional encoder for image compression, illustrating their role in broader ML tasks. An overview of the process is shown in Fig.~\ref{fig:decoder}. While constrained by the scale and limitations of today’s quantum hardware, this study serves as a proof of concept, showcasing how quantum technologies can handle increasingly complex problems. Our work aims to spark curiosity and encourage further exploration of quantum-enhanced approaches in ML, medical imaging and beyond.

\section*{Quantum Annealing}

We focus on the particular implementation of AQC known as QA, in which the problem to be solved is mapped to the minimization of an Ising Hamiltonian, in order to be embeddable in a quantum annealer. FS is also suitable for this technique since it can be formulated as a QUBO problem, which in turn can easily be mapped to an Ising system. QA operates on the fundamental concept of quantum tunneling and the quantum adiabatic theorem to explore the vast solution space of a problem in order to find the optimal solution of the Ising Hamiltonian, which corresponds to a global minimum of some potential function. It is important to note that while QA has shown promise in a variety of fields~\cite{kadowaki1998quantum, heim2015quantum, harris2018phase,king2018observation,albash2021comparing,king2021scaling, abel2022genetic, abel2023string}, it is not a universal quantum computing approach like gate-based quantum computers. Indeed, future gate-based quantum computers are in principle versatile and capable of performing a wide range of computations, while quantum annealers are specialized devices tailored for optimization problems. 

This is not a limitation for our analysis, as our problem can be cast into a form embeddable in a quantum annealer. The QA device we use in this study is developed by D-Wave Systems, featuring specialized hardware that generates and sustains the necessary quantum states. These devices use superconducting qubits and magnetic fields to create controlled quantum environments where the annealing process takes place. In particular, we shall perform our analysis on the \texttt{Advantage\_system4.1} architecture \cite{LantingAQC2017}: this annealer contains 5627 qubits, connected in a \emph{Pegasus} structure, but only has a total of 40279 couplings between them. This topology is depicted in Fig.~\ref{fig:combined_figures}.

The full Hamiltonian in QA comprises an admixture of this Ising {\it problem-Hamiltonian} and a trivial Hamiltonian, called {\it driver Hamiltonian}, for which the ground state is known. The original idea behind QA (and AQC more generally) is to begin in the ground state of the trivial system and adiabatically replace the trivial Hamiltonian with the problem Hamiltonian, while remaining in the ground state throughout. Provided we can remain in the ground state, the final configuration will yield a solution to the problem. In particular, the Hamiltonian takes the form of a generalized Ising model:

\begin{equation}
    \mathcal{H} ~=~ B(s) \, \left( \sum_{ij} J_{ij} \sigma_i^z \sigma_j^z \, + \, \sum_i h_i \sigma_i^z\right) \, + \, A(s)\sum_i\sigma_i^x \, ,
\label{eq:anneal_H}
\end{equation}
where $i$,$j$ label the qubits, $\sigma_i^z$ are the $z$-spin Pauli matrices, and $\sigma_i^x$ are the transverse field components, while the couplings $h_i$ and $J_{ij}$ between the qubits are set and kept constant.
The parameter $s(t)$ (with $t$ being time) is a user-defined control-parameter that can be adjusted, while $A(s)$ and $B(s)$ describe the consequent change in the quantum characteristics of the annealer. As shown in Fig.~\ref{fig:combined_figures}, smaller $s \in [0, 1]$ means larger transverse field parameter $A$ compared to $B$, which induces more ``hopping'' of $\sigma^z$ spins, which overall means a system that is more characteristically ``quantum''.

The anneal schedule increases linearly with time, with $s(0)=0$ and $s(t_f)=1$ (eventually with a pause in between), where $t_f$ is the total annealing time. The network of qubits starts in a global superposition over all possible classical states and, as $s \to 1$, the system localises into a single classical state once a measurement of $\sigma_i^z$ on all sites has been performed. To perform the task of finding a global optimization, the first objective is to encode the problem to be solved into the ``classical'' Ising model Hamiltonian represented by the $B$-terms, which reads
\begin{equation}
\label{eq:classical_P}
    H_P ~=~ \sum_{ij} \,  J_{ij} \,  \mu_i \mu_j \,+\, \sum_i \, h_i \, \mu_i \, ,
\end{equation}
where $P$ stands for ``problem'' and $\mu_i \in \{-1/2, 1/2\}$ are the eigenvalues of the spin operator. In this way, the energetic minimum of the Hamiltonian above would correspond to the desired solution. Then $s$ is adjusted to alter the relative sizes of the parameters $A$, $B$ to perform a so-called anneal, in the hope that the system will end up in the global minimum.

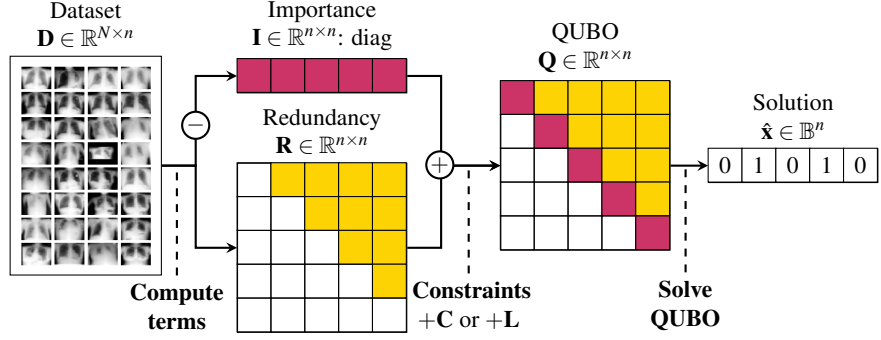
\begin{figure}[!ht]
    \centering
    \begin{tikzpicture}[node distance=1cm, scale=0.9, every node/.style={scale=0.9}]

    % Styles
    \tikzstyle{block} = [rectangle, minimum width=2.5cm, minimum height=0.5cm, text centered, draw=black, fill=purple!30]
    \tikzstyle{matrixblock} = [rectangle, minimum width=2.5cm, minimum height=2.5cm, draw=black]
    \tikzstyle{penaltyblock} = [rectangle, minimum width=4cm, minimum height=0.8cm, draw=black, fill=green!30]
    \tikzstyle{arrow} = [thick,->,>=stealth]
    \tikzstyle{imageblock} = [rectangle, minimum width=2cm, minimum height=3cm, draw=black]

    % Dataset Block
    \node (dataset) [imageblock, label={[align=center]above: Dataset\\$\mathbf{D} \in \mathbb{R}^{N \times n}$}] {\includegraphics[width=2cm,height=3cm]{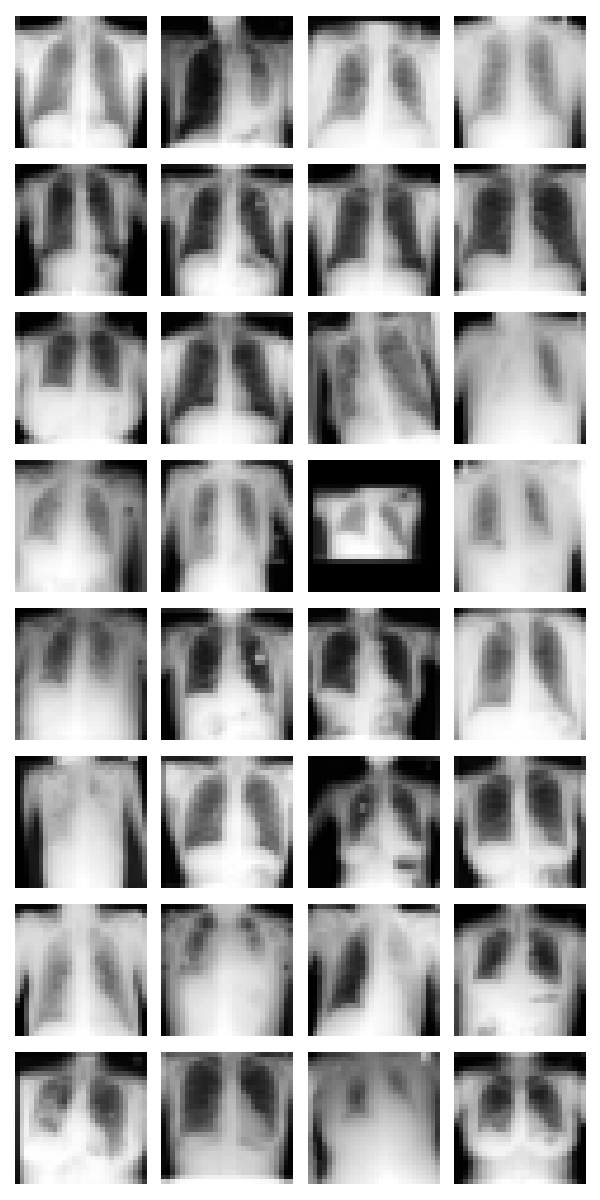}};

    % Importance Block
    \node (importance) [block, above right=-0.5cm and 1.0cm of dataset, label={[align=center]above: Importance\\$\mathbf{I} \in \mathbb{R}^{n \times n}$: diag}] {};

    % Redundancy Block
    \node (redundancy) [matrixblock, below right=-1.5cm and 1.0cm of dataset, label={[align=center]above: Redundancy\\$\mathbf{R} \in \mathbb{R}^{n \times n}$}] {};

    % QUBO Block
    \node (qubo) [matrixblock, right=4.5cm of dataset, label={[align=center]above: QUBO\\$\mathbf{Q} \in \mathbb{R}^{n \times n}$}] {};

    % Solution Block
    \node (solution) [block, right=0.5cm of qubo, label={[align=center]above: Solution\\$\hat{\mathbf{x}} \in \mathbb{B}^{n}$}] {};

    % Grids in Importance (Dark Purple Fill)
    \foreach \x in {0, 0.5, 1, 1.5, 2} {
        \foreach \y in {0} {
            \fill[purple!80] ([xshift=\x cm, yshift=-\y cm]importance.north west) rectangle ++(0.5cm, -0.5cm);
            \draw ([xshift=\x cm, yshift=-\y cm]importance.north west) rectangle ++(0.5cm, -0.5cm);
        }
    }

    % Grids in Redundancy (Upper Triangular with Dark Yellow Fill)
    \foreach \x in {0, 0.5, 1, 1.5, 2} {
        \foreach \y in {0, 0.5, 1, 1.5, 2} {
            \pgfmathparse{\x > \y} % Evaluate condition
            \ifdim\pgfmathresult pt>0pt % Check result
                \fill[yellow!70!orange] ([xshift=\x cm, yshift=-\y cm]redundancy.north west) rectangle ++(0.5cm, -0.5cm);
            \fi
            \draw ([xshift=\x cm, yshift=-\y cm]redundancy.north west) rectangle ++(0.5cm, -0.5cm);
        }
    }

    % Grids in QUBO (Diagonal with Importance Colors, Upper Triangular with Dark Yellow)
    \foreach \x in {0, 0.5, 1, 1.5, 2} {
        \foreach \y in {0, 0.5, 1, 1.5, 2} {
            \pgfmathparse{\x == \y} % Check for diagonal elements
            \ifdim\pgfmathresult pt>0pt
                \fill[purple!80] ([xshift=\x cm, yshift=-\y cm]qubo.north west) rectangle ++(0.5cm, -0.5cm);
            \else
                \pgfmathparse{\x > \y} % Check for upper triangular elements
                \ifdim\pgfmathresult pt>0pt
                    \fill[yellow!70!orange] ([xshift=\x cm, yshift=-\y cm]qubo.north west) rectangle ++(0.5cm, -0.5cm);
                \fi
            \fi
            \draw ([xshift=\x cm, yshift=-\y cm]qubo.north west) rectangle ++(0.5cm, -0.5cm);
        }
    }

    \foreach \x [count=\i from 0] in {0, 0.5, 1, 1.5, 2} {
        \foreach \y in {0} {
            % Draw the cell background explicitly to control color
            \fill[white] ([xshift=\x cm, yshift=-\y cm]solution.north west) rectangle ++(0.5cm, -0.5cm); % Use pink if needed
    
            % Alternate between 0 and 1 based on index, centered
            \ifodd\i
                \node at ([xshift=\x cm + 0.25cm , yshift=-\y cm - 0.25cm]solution.north west) {1};
            \else
                \node at ([xshift=\x cm + 0.25cm, yshift=-\y cm - 0.25cm]solution.north west) {0};
            \fi
    
            % Draw the cell border
            \draw ([xshift=\x cm, yshift=-\y cm]solution.north west) rectangle ++(0.5cm, -0.5cm);
        }
    }

    % Arrows
    \draw [arrow] (dataset.east) -- ++(0.5,0) |- (importance.west);
    \draw [arrow] (dataset.east) -- ++(0.5,0) |- (redundancy.west);
    \draw [arrow] (importance.east) -- ++(0.5,0) |- (qubo.west);
    \draw [arrow] (redundancy.east) -- ++(0.5,0) |- (qubo.west);
    \draw [arrow] (qubo.east) -- (solution.west);

    % Labels for combining operations
    \node at ($(dataset.east)!0.45!(importance.west)+ (0, 0)$) [circle, draw, thick, fill=white, text=black, minimum size=0.4cm, inner sep=0] {$-$};
    \node at ($(redundancy.east)!0.35!(qubo.west) + (0, 0.8)$) [circle, draw, thick, fill=white, text=black, minimum size=0.4cm, inner sep=0] {$+$};

    % Text boxes with arrows
    \node[below=1.7cm of $(dataset.east)!0.2!(importance.west)$, align=center] (text1) {\textbf{Compute} \\ \textbf{terms}};
    \draw[thick, dashed] (text1.north) -- ($(dataset.east)!0.2!(importance.west)+ (0, -0.28)$);

    \node[below=1.84cm of $(importance.east)!0.66!(qubo.west)$, align=center] (text2) {\textbf{ Constraints} \\ $+\mathbf{C}$ or $+\mathbf{L}$};
    \draw[thick, dashed] (text2.north) -- ($(importance.east)!0.66!(qubo.west)+ (0, -0.44)$);

    \node[below=1.42cm of $(qubo.east)!0.4!(solution.west)$, align=center] (text3) {\textbf{Solve} \\ \textbf{QUBO}};
    \draw[thick, dashed] (text3.north) -- ($(qubo.east)!0.4!(solution.west)$);

    \end{tikzpicture}
    \caption{Feature selection pipeline: Images are flattened to compute the importance and redundancy terms, which are combined into the QUBO. The $k$ of $n$ constraint is enforced via a linear penalty or a quadratic constraint (Fig.~\ref{fig:constraint}). Then, the QUBO is solved using classical or quantum solvers.}

    \label{fig:qubo_pipeline}
\end{figure}

\section*{Methods}

\subsection*{Mutual Information Feature Selection QUBO}

The goal for our FS method is to define an optimization objective tailored to a QUBO and complying with the constraints introduced by the current hardware. Consider an image dataset consisting of square, two-dimensional images with width \(W\), paired with their labels. The training dataset can be represented as \(D = \{(\mathbf{X}^i, y^i)\}_{i \in [N]}\), where each \(\mathbf{X}^i \in \mathbb{R}^{W \times W}\) is an image, and \(y^i \in \mathbb{N}\) is its corresponding class label. By flattening the images, the dataset transforms into \(D_f = \{(\mathbf{x}^i, y^i)\}_{i \in [N]}\), where \(\mathbf{x}^i \in \mathbb{R}^n\) represents the \(n\)-dimensional vectorized data, and \(y^i\) remains the class label.  

While the class label \( y^i \) is not used in the subsequent image reconstruction process, it plays a crucial role in feature selection. Specifically, we propose a model-agnostic FS task, where we want to maximize the mutual information (MI) of the features with the class label to determine feature relevance. The selected features are then used the image reconstruction task, independent of \( y^i \). In our simplified example, we will treat the image pixels as features and formulate the pixel selection problem as a QUBO model:

 \begin{equation}
\label{eqn:qubo}
    \min (f_{Q}(\hat{\mathbf{x}})) = \hat{\mathbf{x}}^{T} \mathbf{Q} \hat{\mathbf{x}} = \sum_{i=1}^{n} Q_{i,i} \hat{x}_{i} + \sum_{i=1}^{n} \sum_{j>i}^{n} Q_{i,j} \hat{x}_i \hat{x}_j \, .
\end{equation}

Here, $\hat{x}_{i}$ indicates whether a feature is selected ($1$) or not ($0$), where $\hat{\mathbf{x}} \in \{0,1\}^n$. The linear terms originate from the binary nature of $\hat{x}_{i}^{2} = \hat{x}_{i}$. In turn, our matrix $\mathbf{Q} \in \mathbb{R}^{n \times n}$ describes our optimization problem. In the subsequent steps, we show how we construct the QUBO matrix from the information contained in the image datasets. Intuitively, feature $x_i$ is more likely to be selected if $Q_{i,i}$ is low. Similarly, we can increase the chances of choosing feature $x_{i}$ and feature $x_{j}$ together if the $Q_{i,j}$ term is small. This model closely resembles the form of an Ising problem, and in fact one can translate a QUBO into an Ising problem by mapping  $x_i \rightarrow \mu_i + \frac{1}{2} $, where $\mu_i$ is defined as in Eq.~\eqref{eq:classical_P}.

%\begin{equation}
%    x_i \longrightarrow s_i + \frac{1}{2} \,,
%\end{equation}

In FS, it is common to select features based on MI. Therefore, we closely follow the MI FS QUBO described in~\cite{dwave2020feature, mucke2023feature}. To calculate MI, we estimate the joint and marginal probabilities of the features and labels. For efficient computation with continuous, real-valued features, this requires discretization. We divide each feature dimension into \( B = 20\) bins using quantiles, assigning each feature value to its corresponding bin. Labels, being discrete, do not require binning. This discretization and probability estimation allows for efficient computation of MI between features and labels. Please refer to~\cite{mucke2023feature} for a detailed explanation of this method. From the discretized dataset \( \hat{D} = \{(\mathbf{b}^{i}, y^i)\}_{i \in [N]} \), the empirical joint and marginal probabilities are defined as follows:

\begin{equation}
 \hat{p}_{X, Y}(b, y) = \frac{1}{N} \sum_{j=1}^N \mathbbm{1}(b_j = b \land y_j = y)
\end{equation}

\begin{equation}
    \hat{p}_{X_i, X_j}(b_i, b_j) = \sum_{y} \sum_{b_{k \neq i,j}} \hat{p}_{X, Y}(b, y) \quad\text{and}\quad \hat{p}_{X_i, Y}(b_i, y) = \sum_{b_{k \neq i}} \hat{p}_{X, Y}(b, y)
\end{equation}

\begin{equation}
    \hat{p}_Y(y) = \sum_{b_k} \hat{p}(b, y) \quad\text{and}\quad \hat{p}_{X_i}(b_i) = \sum_{y} \sum_{b_{k \neq i}} \hat{p}_{X, Y}(b, y)
\end{equation}

We initialize our linear QUBO terms with the negative MI of a feature with its class label, which we label as importance:

\begin{equation}
    Q_{i,i} = -I_{i,i} = -I(x_i, y) \approx \sum_{b} \sum_{y} \hat{p}_{X_i, Y}(b, y) \log \frac{\hat{p}_{X_i, Y}(b, y)}{\hat{p}_{X_i}(b) \hat{p}_Y(y)}.
\end{equation}

We want to avoid choosing features that share a lot of information. This is achieved by populating the off diagonal term $Q_{i,j}$ with the MI between feature $x_i$ and $x_j$. A high MI value for $Q_{i,j}$ will increase the energy of the solution that selects both the $x_i$ and $x_j$ features decreasing the probability of it being returned by the annealer.

\begin{equation}
    Q_{i,j}= R_{i,j} = R(x_i, x_j) \approx \sum_{b} \sum_{b'} \hat{p}_{X_i, X_j}(b, b') \log \frac{\hat{p}_{X_i, X_j}(b, b')}{\hat{p}_{X_i}(b) \hat{p}_{X_j}(b')}.
\end{equation}

\begin{figure}[!ht]
    \centering
    \begin{tikzpicture}[node distance=1cm, scale=0.9, every node/.style={scale=0.9}]

    % Styles
    \tikzstyle{matrixblock} = [rectangle, minimum width=2.5cm, minimum height=2.5cm, draw=black]
    \tikzstyle{penaltyblock} = [rectangle, minimum width=4cm, minimum height=0.8cm, draw=black, fill=green!30]
    \tikzstyle{arrow} = [thick,->,>=stealth]
    \tikzstyle{imageblock} = [rectangle, draw=white, inner sep=0]

    % QUBO Block
    \node (qubo) [matrixblock, label={[align=center]above: Thresholded QUBO\\$\mathbf{Q} \in \mathbb{R}^{n \times n}$}] {};
    % Grids in QUBO (Diagonal with Importance Colors, Upper Triangular with Dark Yellow)
    % Grids in Redundancy (Upper Triangular with Dark Yellow Fill)
    \foreach \x in {0, 0.5, 1, 1.5, 2} {
        \foreach \y in {0, 0.5, 1, 1.5, 2} {
            % Map coordinates to integer indices
            \pgfmathsetmacro{\i}{int(round(\x * 2))}
            \pgfmathsetmacro{\j}{int(round(\y * 2))}
            
            % Check for diagonal elements
            \ifnum\i=\j
                \fill[purple!80] ([xshift=\x cm, yshift=-\y cm] qubo.north west)
                    rectangle ++(0.5cm, -0.5cm);
            \else
                \ifnum\i>\j
                    % Specify which cells to fill
                    \pgfmathtruncatemacro{\shouldfill}{
                        (\i==1 && \j==0) || (\i==3 && \j==0) || (\i==2 && \j==1) || (\i==4 && \j==1) || (\i==3 && \j==2) || (\i==4 && \j==2) || (\i==4 && \j==3) ? 1 : 0
                    }
                    \ifnum\shouldfill=1
                        \fill[yellow!70!orange] ([xshift=\x cm, yshift=-\y cm] qubo.north west)
                            rectangle ++(0.5cm, -0.5cm);
                    \fi
                \fi
            \fi
            % Draw the cell border
            \draw ([xshift=\x cm, yshift=-\y cm] qubo.north west)
                rectangle ++(0.5cm, -0.5cm);
        }
    }
    
    % qubo Block
    \node (qubo_function) [imageblock, below=0.1cm of qubo, label={[align=center]below: Hamming weight $w(\hat{\mathbf{x}})$}, label={[align=center]left: $f(\hat{\mathbf{x}})$}] {\includegraphics[width=2.0cm,height=2.0cm]{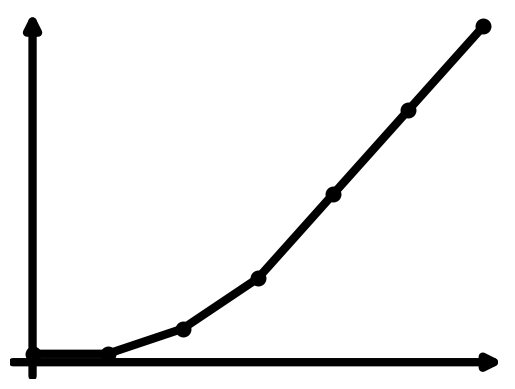}};

    % Quadratic Block
    \node (quadratic_constraint) [matrixblock, above right=0.5cm and 1.5cm of qubo_function, label={[align=center]above: Quadratic constraint\\$\mathbf{C} \in \mathbb{R}^{n \times n}$}] {};

    % Grids in Quadratic (Upper Triangular with Dark Yellow Fill)
    \foreach \x in {0, 0.5, 1, 1.5, 2} {
        \foreach \y in {0, 0.5, 1, 1.5, 2} {
            \pgfmathparse{\x == \y} % Check for diagonal elements
            \ifdim\pgfmathresult pt>0pt
                \fill[purple!80] ([xshift=\x cm, yshift=-\y cm]quadratic_constraint.north west) rectangle ++(0.5cm, -0.5cm);
            \else
                \pgfmathparse{\x > \y} % Check for upper triangular elements
                \ifdim\pgfmathresult pt>0pt
                    \fill[yellow!70!orange] ([xshift=\x cm, yshift=-\y cm]quadratic_constraint.north west) rectangle ++(0.5cm, -0.5cm);
                \fi
            \fi
            \draw ([xshift=\x cm, yshift=-\y cm]quadratic_constraint.north west) rectangle ++(0.5cm, -0.5cm);
        }
    }

    % Linear Block
    \node (linear_constraint) [matrixblock, above right=-2.5cm and 1.5cm of qubo_function, label={[align=center, yshift=-0.05cm]above: Linear constraint\\$\mathbf{L} \in \mathbb{R}^{n \times n}$: diag}] {};

    % Grids in Linear (Upper Triangular with Dark Yellow Fill)
    \foreach \x in {0, 0.5, 1, 1.5, 2} {
        \foreach \y in {0, 0.5, 1, 1.5, 2} {
            \pgfmathparse{\x == \y} % Check for diagonal elements
            \ifdim\pgfmathresult pt>0pt
                \fill[purple!80] ([xshift=\x cm, yshift=-\y cm]linear_constraint.north west) rectangle ++(0.5cm, -0.5cm);
            \fi
            \draw ([xshift=\x cm, yshift=-\y cm]linear_constraint.north west) rectangle ++(0.5cm, -0.5cm);
        }
    }

    % Quadratic Constraint Block
    \node (quadratic_function) [imageblock, right=0.2cm of quadratic_constraint, label={[align=center]below: $w(\hat{\mathbf{x}})$}, label={[align=center, yshift=7.5ex, xshift=4.0ex]left: $c(\hat{\mathbf{x}})$}] {\includegraphics[width=2.0cm,height=2.0cm]{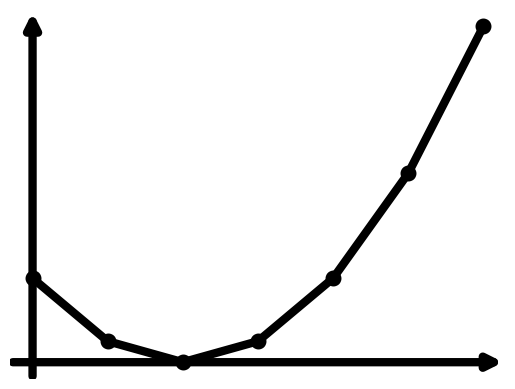}};

    % Linear Constraint Block
    \node (linear_function) [imageblock, right=0.2cm of linear_constraint, label={[align=center]below: $w(\hat{\mathbf{x}})$}, label={[align=center, yshift=7.5ex, xshift=4.0ex]left: $c(\hat{\mathbf{x}})$}] {\includegraphics[width=2.0cm,height=2.0cm]{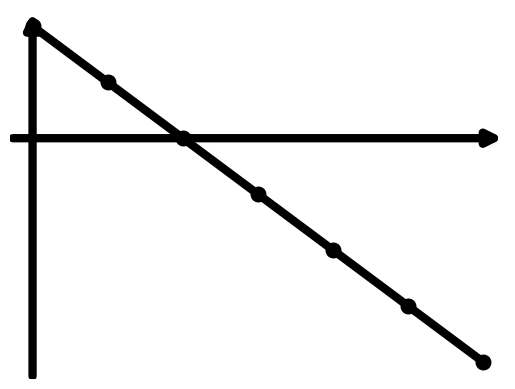}};

    % Quadratic combined Constraint Block
    \node (quadratic_combined_function) [imageblock, right=0.7cm of quadratic_function, label={[align=center]below: $w(\hat{\mathbf{x}})$}, label={[align=center, yshift=7.5ex, xshift=4.0ex]left: $f_{c}(\hat{\mathbf{x}})$}] {\includegraphics[width=2.0cm,height=2.0cm]{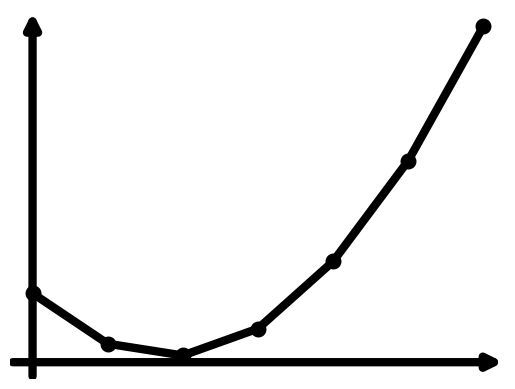}};

    % Linear combined Constraint Block
    \node (linear_combined_function) [imageblock, right=0.7cm of linear_function, label={[align=center]below: $w(\hat{\mathbf{x}})$}, label={[align=center, yshift=7.5ex, xshift=4.0ex]left: $f_{c}(\hat{\mathbf{x}})$}] {\includegraphics[width=2.0cm,height=2.0cm]{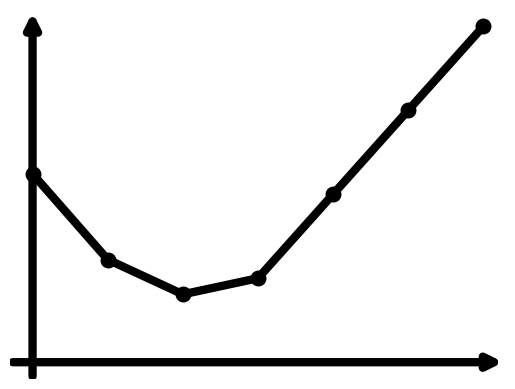}};

    \draw [arrow] (qubo.south east) -- ++(0.75,0) |- (quadratic_constraint.west);
    \draw [arrow] (qubo.south east) -- ++(0.75,0) |- (linear_constraint.west);
    \draw [arrow] (linear_function.east) -- (linear_combined_function.west);
    \draw [arrow] (quadratic_function.east) -- (quadratic_combined_function.west);

    \node at ($(qubo.east)!0.5!(quadratic_constraint.west)+ (0.05, -0.6)$) [circle, draw, thick, fill=white, text=black, minimum size=0.4cm, inner sep=0] {$+$};
    \node at ($(qubo.east)!0.5!(linear_constraint.west) + (0.05, -0.6)$) [circle, draw, thick, fill=white, text=black, minimum size=0.4cm, inner sep=0] {$+$};

    \end{tikzpicture}
    \caption{Illustration of the conventional quadratic constraint to enforce selecting $k$ of $n$ features, which is infeasible due to limited connectivity on the annealer. When dealing with a sparsified QUBO, we propose a linear Ising penalty to enforce the constraint. The QUBO and its constraint are shown, along with a plot of the optimization energy (y-axis) against the Hamming weight (x-axis), indicating how many features are selected.}
    \label{fig:constraint}
\end{figure}
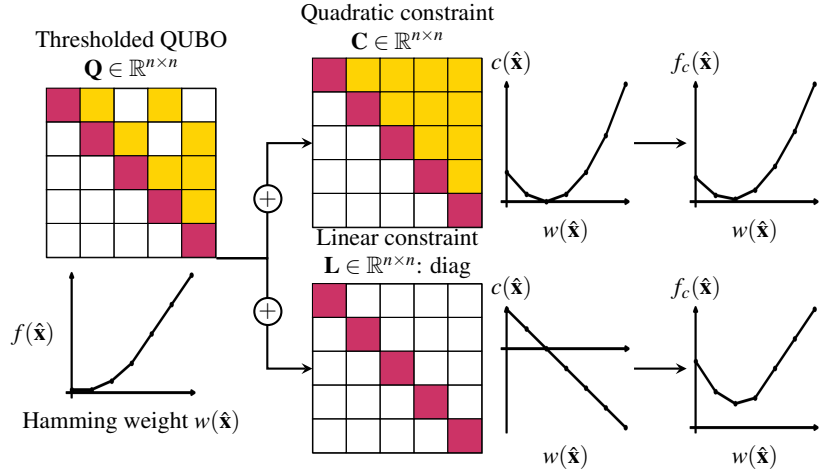

 Finally, we have to enforce our constraint of choosing $k$ of $n$ features. To enforce such constraint, an option is to use a quadratic penalty $\alpha (\sum_{i=1}^{n} \hat{x}_{i} - k)^2$, where $\alpha$ must be tuned to weight the constraint appropriately. The quadratic constraint in the form of a QUBO $\mathbf{C}$ is constructed by $C_{i,i} = 1 - 2k$ and $C_{i,j} = -2n + 2$, resulting in a fully connected problem graph. We construct our FS QUBO problem by additively combining the individual matrices: $\mathbf{Q} = - \mathbf{I} + \mathbf{R} + \alpha \mathbf{C}$. The full procedure of creating the QUBO is depicted in Fig.~\ref{fig:qubo_pipeline} and an illustration of the quadratic constraint is shown in Fig.~\ref{fig:constraint}.

\subsection*{Mapping the Problem to Quantum Hardware}
In the previous section, we demonstrated how to build an MI-based QUBO model for FS. Although this formulation can be run on classical solvers like simulated annealing or tabu-search algorithms, the size and connectivity of the problem graph pose challenges for current quantum annealers. Although modern quantum annealers have a substantial qubit count, their limited connectivity and reliance on chains of qubits to represent logical variables hinder the scalability of highly connected problems, requiring embedding to fit problems onto the QA topology.

To address these challenges, we reduce the intrinsic connectivity of the problem through subsampling and sparsification. Specifically, we down-scale the QUBO by selecting the most informative pixel from local neighborhoods, ensuring that quadratic interactions are retained for the down-scaled features. This significantly reduces the size of the QUBO model while preserving the essential interactions. To further sparsify the QUBO model, we apply a thresholding strategy that eliminates weaker quadratic couplings, ensuring compatibility with the connectivity constraints of the QA hardware. Furthermore, we enforce the $k$ of $n$ constraint using a sparsity-preserving technique, effectively reducing the problem’s complexity and adapting it to hardware limitations.

\subsection*{Linear Ising Penalties}
To limit the connectivity of the QUBO model while still enforcing the constraint, we propose using a linear penalty term, $\left| \sum_{i=1}^{n} \hat{x}_{i} - k \right|$, similar to the approach presented in~\cite{ohzeki2020breaking, mirkarimi2024quantum}. This linear penalty, denoted as $\mathbf{L}$, introduces an offset $\alpha_{l}$ on the diagonal of the QUBO matrix.  The parameter $\alpha_{l}$ is tuned to achieve the desired Hamming weight of the solution vector $\hat{\mathbf{x}}$, where the Hamming weight corresponds to the number of selected features in the solution. The tuning process for $\alpha_{l}$ follows a grid search similar to that used for weighting the quadratic constraint. When constructing the QUBO we substitute $\mathbf{C}$ for $\mathbf{L}$: $\mathbf{Q} = - \mathbf{I} + \mathbf{R} + \alpha_{l} \mathbf{L}$. An overview and visual comparison of the constraint creation process for the QUBO is provided in Fig.~\ref{fig:constraint}. While the linear penalty constraint may be less effective for certain problem instances, our experiments show it consistently enforces the desired behavior.

\subsection*{Reconstruction Decoder}
Our experiment focuses on reconstructing images from the selected subset of pixels. Once the features are extracted for each dataset, we train a convolutional decoder to reconstruct the original image from the selected pixels. An illustration of the procedure in displayed in Fig.~\ref{fig:decoder}. The reconstruction network consists of a linear layer followed by two two-dimensional transposed convolutional layers, each followed by a ReLU activation function and then a sigmoid activation function. 

\subsection*{Datasets}
For our experiments, we used the MedMNIST dataset~\cite{yang2023medmnist}, which contains 18 standardized datasets used for biomedical image classification. The collections compromise 12 two-dimensional and 6 three-dimensional datasets. The collection contains data scales from a few hundred to 100,000 and binary and multi-class classification tasks. In particular, the dataset collection should facilitate light-weight machine learning research in medical imaging without directly facing clinical challenges. Due to the size restrictions of our QA device, we only analyze the two-dimensional grayscale image datasets. We note that ChestMNIST, OCTMNIST, PneumoniaMNIST and BreastMNIST consist of images that are semi-registered. This plays an important role when discussing pixels as feature descriptors. The datasets analyzed are summarized in Table~\ref{tab:dataset}.

\begin{table}[b!]
    \centering
    \begin{tabular}{lcccc}
        \hline
        Dataset $(28 \times 28)$ & Modality & Tasks (Classes) & \# Samples & Train/Test \\ \hline
        ChestMNIST & Chest X-Ray & Binary (2) & 112,120 & 78,468/22,433 \\
        OCTMNIST & Retinal OCT & Multi (4) & 109,309 & 97,477/1,000\\
        PneumoniaMNIST & Chest X-Ray & Binary (2) & 5,856 & 4,708/624\\
        BreastMNIST & Breast Ultrasound & Binary (2) & 780 & 546/156\\
        TissueMNIST & Microscope & Multi (8) & 236,386 & 165,466/47,280 \\
        OrganAMNIST & Abdominal CT & Multi (11) & 58,830 & 34,561/17,778 \\ \hline
    \end{tabular}
    \caption{MedMNIST v2 2D datasets used for the feature selection experiments in this manuscript. Table adapted from~\cite{yang2023medmnist}.}
    \label{tab:dataset}
\end{table}

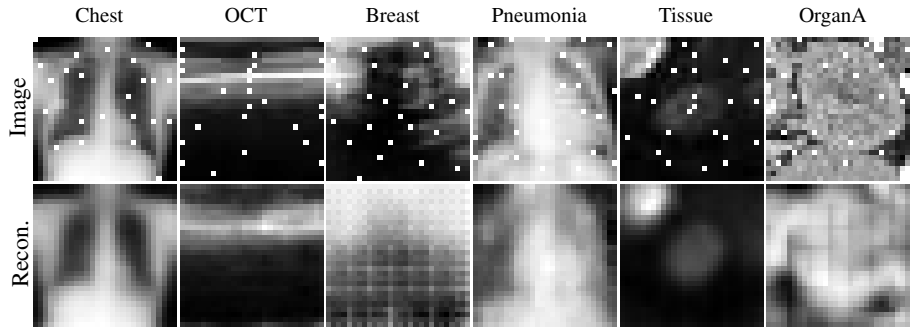
\begin{figure}[!ht]
    \centering
    \begin{tikzpicture}
    
        % Parameters
        \def\imgsize{1.9cm}  % Image size (width and height)
        \def\spacing{0.05cm}  % Spacing between images
        \def\captionfont{\scriptsize}   % Font size for captions
        
        % Top row (Overlay images)
        \node[anchor=north west] at (0, 0) {\includegraphics[width=\imgsize, height=\imgsize]{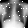}};
        \node[anchor=north west] at (\imgsize + \spacing, 0) {\includegraphics[width=\imgsize, height=\imgsize]{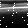}};
        \node[anchor=north west] at (2 * \imgsize + 2 * \spacing, 0) {\includegraphics[width=\imgsize, height=\imgsize]{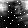}};
        \node[anchor=north west] at (3 * \imgsize + 3 * \spacing, 0) {\includegraphics[width=\imgsize, height=\imgsize]{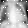}};
        \node[anchor=north west] at (4 * \imgsize + 4 * \spacing, 0) {\includegraphics[width=\imgsize, height=\imgsize]{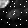}};
        \node[anchor=north west] at (5 * \imgsize + 5 * \spacing, 0) {\includegraphics[width=\imgsize, height=\imgsize]{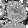}};
        
        % Bottom row (Reconstructed images)
        \node[anchor=north west] at (0, -\imgsize - \spacing) {\includegraphics[width=\imgsize, height=\imgsize]{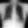}};
        \node[anchor=north west] at (\imgsize + \spacing, -\imgsize - \spacing) {\includegraphics[width=\imgsize, height=\imgsize]{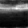}};
        \node[anchor=north west] at (2 * \imgsize + 2 * \spacing, -\imgsize - \spacing) {\includegraphics[width=\imgsize, height=\imgsize]{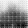}};
        \node[anchor=north west] at (3 * \imgsize + 3 * \spacing, -\imgsize - \spacing) {\includegraphics[width=\imgsize, height=\imgsize]{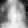}};
        \node[anchor=north west] at (4 * \imgsize + 4 * \spacing, -\imgsize - \spacing) {\includegraphics[width=\imgsize, height=\imgsize]{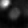}};
        \node[anchor=north west] at (5 * \imgsize + 5 * \spacing, -\imgsize - \spacing) {\includegraphics[width=\imgsize, height=\imgsize]{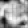}};
        
        % Rotated row captions on the left
        \node[anchor=center, rotate=90] at (-\spacing, -\imgsize / 2 - 2*\spacing) {\small Image};
        \node[anchor=center, rotate=90] at (-\spacing, -\imgsize - \spacing - \imgsize / 2 - 2*\spacing) {\small Recon.};
        
        % Column captions on the top
        \node[anchor=south] at (\imgsize / 2 + \spacing, -\spacing) {\footnotesize Chest};
        \node[anchor=south] at (\imgsize + \spacing + \imgsize / 2 + \spacing, -\spacing) {\footnotesize OCT};
        \node[anchor=south] at (2 * \imgsize + 2 * \spacing + \imgsize / 2 + \spacing, -\spacing) {\footnotesize Breast};
        \node[anchor=south] at (3 * \imgsize + 3 * \spacing + \imgsize / 2 + \spacing, -\spacing) {\footnotesize Pneumonia};
        \node[anchor=south] at (4 * \imgsize + 4 * \spacing + \imgsize / 2 + \spacing, -\spacing) {\footnotesize Tissue};
        \node[anchor=south] at (5 * \imgsize + 5 * \spacing + \imgsize / 2 + \spacing, -\spacing -0.05cm) {\footnotesize OrganA};
    \end{tikzpicture}
    \caption{Visual comparison of images from the test set, overlayed with the selected pixels (top row) and the decoder reconstructed images from the selected pixels.}
    \label{fig:visual_comparison}
\end{figure}

\section*{Results}

\begin{table}[b!]
\centering
\begin{tabular}{ccccc:c}\hline
Dataset & Random $\times 10^{-3}$ & Sampled$\times 10^{-3}$ & qbsolv$\times 10^{-3}$ & QA $\times 10^{-3}$ & AE$\times 10^{-3}$\\\hline
ChestMNIST & $6.7 \pm 0.6$ & $6.0 \pm 0.0$ & $5.3 \pm 0.2$ & $5.5 \pm 0.2$ & $2.3 \pm 0.0$\\
OCTMNIST & $6.7 \pm 1.8$ & $5.9 \pm 0.1$ & $4.7 \pm 0.1$ & $5.3 \pm 0.5$ & $1.4 \pm 0.1$\\
BreastMNIST & $22.3 \pm 3.5$ & $23.5 \pm 1.3$ & $21.8 \pm 3.0$ & $21.2 \pm 2.6$ & $14.8 \pm 1.7$\\
PneumoniaMNIST & $6.9 \pm 0.5$ & $5.9 \pm 0.4$ &$5.5 \pm 0.1$ & $5.3 \pm 0.2$ & $ 3.1 \pm 0.1$\\
TissueMNIST & $3.6 \pm 0.2$ & $3.1 \pm 0.0$ & $3.3 \pm 0.1$ & $3.0 \pm 0.1$ & $1.6 \pm 0.0$\\
OrganAMNIST & $39.4 \pm 0.9$ & $37.2 \pm 0.3$ & $40.3 \pm 0.4$ & $37.4\pm 0.5$ & $24.7 \pm 0.2$\\\hline
\end{tabular}
\caption{Test set MSE for different pixel selection methods, reported as mean~$\pm$~standard deviation over five independent runs.}
\label{tab:results_reconstruction}
\end{table}

\subsection*{Reconstruction Experiment}
We compare different FS methods for selecting subsets of features in our reconstruction experiments. The FS methods evaluated include: (1) a \textbf{random sampling approach} that selects $k$ features at random, (2) a \textbf{subsampled approach} that evenly spreads the pixels across the image in a subsampled grid fashion, (3) the \textbf{full QUBO formulation} solved using the classical tabu-search solver \texttt{qbsolv}~\cite{qbsolv2017dwave}, (4) our proposed \textbf{quantum annealing (QA) method}, and (5) a \textbf{learned feature extractor}, specifically an autoencoder (AE) with a latent dimension of $k=25$, whose encoder resembles the presented decoder architecture.  

Beyond these approaches, numerous FS techniques exist, including \textbf{random forest-based selection}~\cite{menze2009comparison} and \textbf{simulated annealing-based selection}~\cite{mafarja2017hybrid}. While these were not explicitly compared in our experiments, they represent alternative strategies for feature selection that may be explored in future work.

Feature sets of size $k = 25$ were used to train a convolutional encoder that takes the selected pixels as input to reconstruct the original image. The encoder was trained using the Adam optimizer with a learning rate of 0.001 for 20 epochs, minimizing mean squared error (MSE) loss. Reconstruction performance was validated first on the MNIST dataset, showing results consistent with those reported by Muecke et al.~\cite{mucke2023feature}. The experiments were then extended to the MedMNIST dataset. Each selection and training process was repeated five times to calculate the mean and standard deviation. The reconstruction results, expressed in terms of MSE on the test set, are summarized in Table~\ref{tab:results_reconstruction}. Visual examples of reconstructed images for each dataset using QA-selected pixels are shown in Fig.~\ref{fig:visual_comparison}.

\subsection*{Simulation and Hardware Experiments}
In the classical simulation setup, we performed FS on the full QUBO of size $784 \times 784$, incorporating a quadratic constraint. The QUBO was solved using the \texttt{qbsolv} algorithm, which partitions the problem into smaller subproblems. From the generated sampleset, the solution with the lowest energy was selected. The runtime of \texttt{qbsolv} on the full size QUBO was $5.6 \pm 0.56 \, \mathrm{s} $.

To validate the solutions on real quantum hardware, we executed the experiments on the D-Wave Advantage 4.1 system through Leap\textsuperscript{TM} using the associated Ocean Python API~\cite{dwave2023leap}. As discussed above, we used a series of steps to reduce the size and connectivity of the QUBO model. The original QUBO was defined on a $28 \times 28$ pixel grid, resulting in a size of $784 \times 784$. To downscale this, we used a subsampling strategy, selecting the pixel with the highest MI from each $2 \times 2$ neighborhood in the image. This reduced the QUBO size to $196 \times 196$. Despite this reduction, the connectivity of the QUBO still exceeded the hardware’s constraints. To address this, we applied a thresholding technique that removed weaker quadratic couplings, retaining 2000 couplings in the final problem representation. This sparsification step aligned with an average chain length of 3 and used around 800 qubits on the QA hardware. The $k$ of $n$ constraint was enforced using a sparsity-preserving linear penalty, ensuring compatibility with the hardware while preserving the structure of the problem. The subsampled, thresholded, and linear Ising-constraint-enforced QUBO was mapped to the annealing hardware, with the annealing time set to $t_f = 20 \, \mu s$. We performed 1000 reads to form a sampleset and selected the solution with the lowest energy. The runtime on the quantum annealer was $260.84 \pm 12.64 \, \mathrm{ms}$. Overhead times, such as queueing the problem to the quantum annealer and preparing the QUBO, were not included, as these are shared with the \texttt{qbsolv} workflow.    

\section*{Discussion}
In this work, we presented a method to encode a feature selection (FS) problem that can be implemented on commercially available quantum computing hardware. Our approach focuses on selecting the $k$ most important features, as measured by mutual information (MI), from six lightweight medical image datasets. We evaluated the selected features by training a convolutional reconstruction decoder and measuring the MSE of reconstructed samples compared to ground truth test set images.

To address the limitations of quantum hardware connectivity, we enforced a linear penalty to reduce the connectivity of the problem graph. This allowed us to generate a subsampled, thresholded QUBO formulation, that reduces the connectivity and computational complexity of the problem. Despite these simplifications, the solutions obtained on the quantum hardware were quantitatively (see Table~\ref{tab:results_reconstruction}) and qualitatively (see Fig.~\ref{fig:visual_comparison}) comparable to those derived from a simulated solver operating on the complete problem description. This demonstrates that our method effectively balances the trade-off between hardware limitations and solution quality. The performance of our quantum-based FS approach was comparable to that achieved using a classical solver.

Our experiments showed that the QUBO-based FS method identified plausible features for training the reconstruction encoder. However, the effectiveness of FS was dataset-dependent. For medical imaging datasets with relatively aligned images, such as ChestMNIST, PneumoniaMNIST, BreastMNIST, and OCTMNIST, the method performed well, resulting in meaningful feature subsets that supported accurate reconstructions. In contrast, for datasets where the image content was misaligned or highly heterogeneous, such as cell images, or multi-organ images, the selected features did not outperform simple subsampling. This is consistent with existing knowledge that localized pixel features are suboptimal for these tasks, as they fail to capture global spatial or contextual information. This limitation highlights why neural networks, which excel at learning hierarchical and spatially invariant representations, are particularly effective for such tasks.
It is important to note that the objective of this study was not to propose a new feature extractor but rather to select a subset of features from a large set that adequately describes the data distribution. Learned feature descriptors, such as those produced by neural networks, offer unparalleled performance due to their ability to generalize across diverse datasets, they often lack interpretability. In contrast, our QUBO-based approach provides interpretable FS grounded in statistical measures such as MI and redundancy, offering insights into the most relevant features for specific tasks.

Integrating learned feature representations with interpretable FS methods could present a powerful hybrid approach. For instance, features extracted from foundation models or other pre-trained neural networks could be used as inputs to our QUBO-based framework, combining the strengths of deep learning with the interpretability and sparsity benefits of quantum-inspired FS. This would enable both effective and interpretable solutions, bridging the gap between data-driven learning and human-comprehensible FS. Additionally, future work could investigate the incorporation of more advanced quantum algorithms or optimizing hardware configurations to further improve scalability and performance.

\section*{Conclusion}
In this work, we provided an introduction to using quantum annealers for feature selection. We presented a method to perform feature selection on currently available quantum annealing hardware and applied it on light-weight medical imaging datasets. The method outperforms other simple feature selection techniques, but cannot compete against trainable deep learning based feature extractors. By leveraging an adapted QUBO formulation with thresholding, subsampling and hardware optimized constraints, we demonstrated how quantum hardware can be used effectively despite its current limitations. This work highlights potentials of quantum feature selection as a foundation for future explorations in interpretable and scalable feature selection methodologies, possibly combining deep learning based feature extractors and quantum based feature selection.

\section*{Acknowledgements}

This study was partially funded by the exchange visit program of the EPSRC International Network on Quantum Annealing (EP/W027003/1).

\section*{Author contributions statement}

M.A.N, L.A.N., and B.C. conceived the experiments,  M.A.N. and B.C. conducted the experiments, M.A.N, L.A.N., B.C., P.A.W. and A.K.M. analysed the results. All authors reviewed the manuscript. 

\section*{Competing interests}
The authors have no competing interests to declare that are relevant to the content of this article.
%To include, in this order: \textbf{Accession codes} (where applicable); \textbf{Competing interests} (mandatory statement). 

%The corresponding author is responsible for submitting a \href{http://www.nature.com/srep/policies/index.html#competing}{competing interests statement} on behalf of all authors of the paper. This statement must be included in the submitted article file.

\end{document}